\documentclass[11pt]{article}
\usepackage{graphicx}
\newcommand{\be}{\begin{equation}}
\newcommand{\ee}{\end{equation}}

\newcommand{\bea}{\begin{eqnarray}}
\newcommand{\eea}{\end{eqnarray}}
\newcommand{\mbold}[1]{\mbox{\boldmath$#1$}}
\newcommand{\bnabla}{\mbold{\nabla}}
\title{Semiclassical approximation to the Hartree-Fock method in
finite nuclei}

\author{K.A Gridnev and V.B.Soubbotin \\ {\small  Nuclear Physics
Department, Physical Research Institute},\\ {\small St.Petersburg
University, St.Petersburg, Russia} \\
{X.Vi\~nas and M.Centelles}\\ {\small Departament d'Estructura i
Constituents de la Mat\`eria, Facultat de F\'{\i}sica,}
\\{\small Universitat de Barcelona, Diagonal 647 E-08028 Barcelona,
Spain }}

\date{}

\begin{document}

\maketitle

\begin{abstract}
In this paper we review the semiclassical extended Thomas--Fermi
theory for describing the ground-state properties of nuclei. The
binding energies calculated in this approach do not contain shell
effects and, in this sense, they are analogous to those obtained from
the mass formula. We discuss some techniques for incorporating the
shell effects which are missing in the semiclassical calculation,
such as the so-called expectation value method and the Kohn-Sham
scheme. We present numerical applications for effective zero-range
Skyrme forces and finite-range Gogny forces.
\end{abstract}

\pagebreak

\section{Introduction}

To obtain the ground-state energy and the particle density of a set of
interacting nucleons is one of the most important problems in nuclear
physics. This is a complicated many-body problem if realistic
nucleon-nucleon interactions are used. To overcome the difficulties
effective nucleon-nucleon forces and approximated schemes have been
proposed. One of the most outstanding approaches is the Hartree--Fock
(HF) method, which consists in replacing the many-body wave function by
a Slater determinant of single-particle wave functions. These
wave functions are obtained self-consistently from the mean field
produced by the nucleons.

When used together with effective density-dependent
nucleon-nucleon forces, like for
example the Skyrme \cite{Sky}, Gogny \cite{Gog} or M3Y
\cite{M3Y} interactions, the HF method is a very powerful tool to
carry out in a simple
way accurate nuclear structure calculations. This density-dependent HF
(DDHF) approach yields binding energies and root-mean square radii in
very
good agreement with experiment. These simple forces are also able
to describe dynamical phenomena such as excited nuclei properties,
nuclear excitation spectra, nucleon-nucleus optical potential and
low energy heavy-ion scattering (see for example \cite{Li} for a
review for Skyrme forces).

Another different approach within the mean field
theory is the so-called density functional theory
(DFT). The basic idea of DFT is that the ground-state energy of a
system of interacting fermions can be expressed by an integral over
the whole space of an energy density which depends only on the
ground-state local density $\rho(\mbold{R})$:
\be
E [\rho] = \int \varepsilon [\rho] d \mbold{R} .
\label{eq1} \ee
The theoretical justification of DFT is provided by the
Hohenberg--Kohn theorem \cite{HK}. It states that the exact
non-degenerate ground-state energy of a correlated electron system is
a functional of
the local density $\rho(\mbold{R})$ and that this functional has its
variational minimum when evaluated with the exact ground-state
density:
\be
\frac{\delta}{\delta \rho} \bigg[ E[\rho] - \mu \int \rho(\mbold{R})
d \mbold{R} \bigg]=0 ,
\label{eq2} \ee
where $\mu$ is the Lagrange multiplier for ensuring the right number
of particles.

Unfortunately, $\varepsilon[\rho]$ is not exactly known for a finite
interacting fermion system and consequently approximations are in
order. It is useful to break up the energy of the system in several
pieces by writing 
\be
E[\rho] = T[\rho] + \frac{1}{2} \int V_H[\rho] \rho(\mbold{R})
d \mbold{R} + E_{ex}[\rho] ,
\label{eq3} \ee
where $T[\rho]$ is the kinetic energy of a system
of non-interacting fermions of density $\rho$ and $V_H[\rho]$ is the
direct (Hartree) potential given by
\be
V_H[\rho] = \int v(\mbold{R},\mbold{R'}) \rho(\mbold{R'}) 
d \mbold{R'} ,  
\label{eq4} \ee
with $v(\mbold{R},\mbold{R'})$ the effective
nucleon-nucleon interaction. The last term in eq. (\ref{eq3}) is the
so-called exchange-correlation energy that contains the exchange
energy as well as contributions of the correlations due to the fact
that the exact wave function is not a Slater determinant.

The kinetic energy functional $T[\rho]$ is not known exactly either.
Kohn and Sham (KS) \cite{KS} proposed to
write the local density in terms of trial single-particle
wave functions $\phi_i(\mbold{R})$ as
\be
\rho(\mbold{R}) = \sum^A_{i=1} {\vert \phi_i (\mbold{R}) \vert}^2 ,
\label{eq5} \ee
furthermore assuming that the kinetic energy density
functional can also be expressed through the same trial
single-particle wave functions:
\be
\frac{\hbar^2}{2m} \tau(\mbold{R}) =
\frac{\hbar^2}{2m} \sum^A_{i=1}  
{\vert \bnabla \phi_i (\mbold{R}) \vert}^2 .
\label{eq6} \ee
With the help of eqs. (\ref{eq5})-(\ref{eq6}), the variation with
respect to $\rho$ in eq. (\ref{eq2}) can be easily carried out to
obtain
\be
\big \{ - \frac{\hbar^2}{2 m} \Delta + V_H(\mbold{R})
+ \frac{\delta E_{ex}}{\delta \rho} \big \} \phi_i(\mbold{R})
= \varepsilon_i \phi_i(\mbold{R}) ,
\label{eq7} \ee
which are known as Kohn--Sham equations.
These equations are similar to the HF ones, although the KS
potential 
\be
V_{KS}(\mbold{R}) = V_H(\mbold{R})
+ \frac{\delta E_{ex}}{\delta \rho}
\label{eq8} \ee
is local as compared with the, in general, non-local HF
potential. The difference between the exact kinetic energy density
functional and the approximated functional given by eq.\ (\ref{eq6})
is included in the exchange-correlation term $E_{ex}$.

Although the KS approach is exact if one deals with the exact
$E_{ex}$, this theory is not free of certain difficulties. First, the
physical interpretation of the KS orbitals is not clear. In view of
eq. (\ref{eq7}) they are often considered as a HF wave function.
However,
there is no real justification for it because no assumption is done
about representing the many-body wave function by a Slater
determinant. On the other hand, the physical meaning of the
eigenvalues $\varepsilon_i$ of eq. (\ref{eq7}) is not clear either
because the Koopmans' theorem, valid in the HF approach, does not
apply in the KS scheme. It should also be pointed out that the
kinetic energy density, eq. (\ref{eq6}), is only an approach to the
exact one because the existence of a set of
single-particle wave functions obeying simultaneously eqs. (\ref{eq5})
and (\ref{eq6}) has not been proved \cite{PS}.

Alternatively, another approximation consists in writing the kinetic
energy density functional explicitly in terms of the local density
and its gradients. In such a case the variational equation (\ref{eq2})
allows one to find $\rho(\mbold{R})$ directly, avoiding the task of
solving
the KS equations. The simplest case is just the Thomas-Fermi (TF)
approximation, where the kinetic energy density is written as
\be
T[\rho] = \frac{\hbar^2}{2m} \frac{3}{5} ({\frac{3 \pi^2}{2}})^{2/3}
\rho^{5/3} ,
\label{eq9} \ee
if a degeneracy 4 is assumed.

The TF approach is exact at the quantal level for an uniform system
and therefore does not contain shell effects. When applied to finite
nuclei, it implies that in the neighbourhood of a point $\mbold{R}$
the nucleus behaves as a piece of nuclear matter of density
$\rho(\mbold{R})$. Thus, for a finite nucleus the TF energy represents
an average energy and it is similar in spirit to the one obtained with
the mass
formula derived within the liquid drop or droplet models \cite{MS}.
The success of these semiclassical approaches lies on the fact that
the quantal corrections (shell effects) are small as compared with the
part of the energy that varies slowly with the number of particles
$A$, the one which is provided by the mass formula or by the TF
method.

From a theoretical point of view, the perturbative treatment of the
shell correction energy in finite Fermi systems is based on the
so-called Strutinsky energy theorem \cite{Str}. It states that the
total quantal energy can be split in two parts:
\be
E = \tilde{E}  + \delta E .
\label{eq10} \ee
The largest part $\tilde{E}$ varies smoothly with the number of
particles $A$. It can be calculated in a way similar to the exact
energy $E$, but using the smoothed density matrix (equivalent to
the semiclassical one) obtained with Strutinsky averaged (SA)
occupation numbers instead of the quantal density matrix
\cite{Brack}. The shell correction $\delta
E$ has a pure quantal origin and its behaviour is not smooth at all;
it is much smaller than $\tilde{E}$, although it can become important
in some cases like in low-energy nuclear fission.

Brack and Quentin \cite{BQ} have carried out extensive HF Strutinsky
calculations with Skyrme forces. From these calculations it can be
seen that the properties of $\tilde{E}$ are like those of the
semiclassical liquid droplet model. They have also shown that the
shell effects can be perturbatively added to the self-consistent
smooth quantities. These facts suggest that it could be possible to
replace the microscopic Strutinsky smooth quantities, which are
rather difficult to handle in practice, by a much simpler
semiclassical calculation of them.

It should be pointed out that the pure TF approximation is
not well suited for the variational calculation of eq. (\ref{eq2}).
Corrections to the kinetic energy density that take into account the
finite size of the nucleus have to be included, as for instance the
well-known Weizs\"acker correction:
\be
T_{W}[\rho] = \frac{\hbar^2}{72 m} \frac{(\bnabla{\rho})^2}{\rho} .
\label{eq11} \ee
In a more systematic way, the extended Thomas-Fermi (ETF)
method, which includes up to $\hbar^4$-order corrections to the
kinetic energy and spin-orbit densities through the particle
density and its gradients up to fourth order, has been developed in
the past \cite{Gram,Brack}. These $\hbar^4$-order functionals used in
conjunction with Skyrme forces, lead to fourth order and highly
non-linear
equations for the particle density $\rho(\mbold{R})$ that can be
solved self-consistently \cite{Cen90}. For Skyrme forces the
binding energies for finite nuclei obtained including these
$\hbar^4$-contributions are close to the Strutinsky results
\cite{Brack} although some differences persist due to the
approximations used for obtaining the $T[\rho]$ functional in the
ETF approach \cite{Cen98}.
Very recently \cite{SV}, the ETF approach has been employed to
obtain the semiclassical density matrix up to $\hbar^2$ order in
coordinate space for the case of a non-local HF potential for
finite range effective nucleon-nucleon forces.

This work is devoted to the discussion of semiclassical approaches to
the HF method
and the comparison with full quantal results for zero-range and finite
range nucleon-nucleon effective forces.
The paper is organized as follows. In Section 2 we briefly present the
derivation of the semiclassical ETF HF energy in the case of a finite
range effective interaction. Section 3 is devoted to the study of the
ETF energies using Skyrme forces with recently presented
parametrizations \cite{Cha} which are able to describe nuclei far from
the stability line and a way for including shell
effects is discussed. In Section 4 we perform ETF and KS
calculations using the Gogny force. The conclusions are laid
in the last Section.

\section{Hartree-Fock energy in the Extended Thomas-Fermi
approximation}

In this section we derive the ETF HF energy for a non-local potential
following closely the method presented in Refs. \cite{Cen98,SV}.
We start from the quantal one-body Hamiltonian which for
each kind of particle reads
\be
\hat{H} = - \frac{\hbar^2}{2m} \Delta + V^{H}(\mbold{r},\mbold{r'})
\delta(\mbold{r} - \mbold{r'}) + V^{F}(\mbold{r},\mbold{r'})
- i \mbold{W}(\mbold{r},\mbold{r'}) (\bnabla \times \mbold{\sigma})
\delta(\mbold{r} - \mbold{r'}) ,
\label{eq12} \ee
where the last term is the spin-orbit Hamiltonian. The corresponding
HF energy for an uncharged nucleus can be written as
\bea
E_{HF} &=& \sum_q \int d \mbold{R} \bigg[ \frac{\hbar^2}{2m}
\tau(\mbold{R}) +
\frac{1}{2} \rho(\mbold{R}) V_{dir} (\mbold{R})  + \frac{1}{2}  \int d
\mbold{s} V_{ex}(\mbold{R},s) \rho (\mbold{R} + \frac {\mbold{s}}{2},
\mbold{R} - \frac {\mbold{s}}{2}) \nonumber \\ &+&
\mbold{J}(\mbold{R}) \mbold{W}(\mbold{R}) \bigg]_q,
\label{eq13} \eea
where the subindex $q$ refers to each kind of nucleon.
In eq. (\ref{eq13}) $\mbold{R}=(\mbold{r}+\mbold{r'})/2$ and
$\mbold{s}=\mbold{r}-\mbold{r'}$ are the center of mass and
relative coordinates, and the direct ($V^{H}$) and exchange
($V^{F}$) parts of the HF potential are given by
\be
V^{H}(\mbold{R}) = \int d \mbold{s} v(\mbold{s})
\rho(\mbold{R},\mbold{s})
\label{eq14} \ee
and
\be
V^{F}(\mbold{R}) = -v(\mbold{s}) \rho(\mbold{R}+\frac{\mbold{s}}{2},
\mbold{R}-\frac{\mbold{s}}{2}) .
\label{eq15} \ee
where for the sake of simplicity we use a simple Wigner force.

In terms of the one-body density matrix
$\rho(\mbold{R}+\frac{\mbold{s}}{2},
\mbold{R}-\frac{\mbold{s}}{2}) = \sum^{A}_{i=1} \phi^{*}_i (\mbold{r})
\phi_i (\mbold{r'})$, the particle, kinetic energy and spin-orbit
densities read
\be
\rho(\mbold{R}) = \rho(\mbold{R},\mbold{s}) \vert_{s=0}
\label{eq16} \ee
\be
\tau(\mbold{R}) = \bigg(\frac{1}{4} \Delta_R - \Delta_s \bigg)
\rho(\mbold{R},\mbold{s}) \vert_{s=0}
\label{eq17} \ee
\be
\mbold{J}(\mbold{R}) = -i \bigg[ \mbold{\sigma} \times
(\frac{1}{2}
\bnabla_R + \bnabla_{s}) \bigg] \rho(\mbold{R},\mbold{s})
\vert_{s=0} .
\label{eq18} \ee
The HF ETF energy is obtained from eq. (\ref{eq13}) replacing the
quantal
densities by their corresponding ETF values. The ETF HF energy can be
written as a functional of the local density only if the density
matrix is expressed in terms of $\rho(\mbold{R})$ and its gradients.

The simplest semiclassical approach to the density matrix
corresponds to the so-called Slater or TF approximation where
\be
\rho(\mbold{R},s) = \frac{3 j_1(k_F s)}{k_F s} \rho .
\label{eq19} \ee
In this equation $j_1(k_F s)$ is the $l=1$ spherical Bessel function
and $k_F$ the
local Fermi momentum, which for each kind of nucleon is related with
the local density by
\be
k_F = (3 \pi^2 \rho)^{1/3} .
\label{eq20} \ee
At this TF level there is no semiclassical spin-orbit contribution. On
the other hand, as it has been pointed out in previous literature
the Slater approach to the density matrix cannot be very
accurate for describing the exchange part specially if the non-local
effects are important \cite{NV,SV}.

Corrections to the one-body density matrix that take into account
finite-size effects have been considered in the past \cite{NV,CB}.
However, we will use here the recently developed ETF approach to the
one-body density matrix \cite{SV}.
Usually, the semiclassical methods of TF type are based on the
Wigner-Kirkwood $\hbar$ expansion of the density matrix \cite{WK}.
This expansion can be obtained in several ways. The partition function
approach of Bhaduri \cite{Bha}, the Kirzhnits expansion \cite{Kir},
the algebraic method of Grammaticos and Voros \cite{Gram} or the direct
expansion of the density matrix \cite{Ring} are some examples. We will
use the latter way that has been applied in the case of non-local
potentials in refs. \cite{Cen98,SV}.

The Wigner transform of the quantal single-particle Hamiltonian
(\ref{eq12}) is given by
\be
H_W = \frac{\hbar^2 k^2}{2m} + V^{H}(\mbold{R}) + V^{F}(\mbold{R},k)
- (\mbold{W}(\mbold{R}) \times \mbold{k}) \mbold{\sigma} = \tilde{H}_W
+ H_{so}.
\label{eq21} \ee
In this equation $V^{H}(\mbold{R})$ and $V^{F}(\mbold{R},k)$ are the
Wigner transform of the direct and exchange parts of the HF potential
and are given by
\be
V^{H}(\mbold{R}) = \int d \mbold{R'} v(\mbold{R},\mbold{R'}) \rho
(\mbold{R'})
\label{eq22} \ee
and
\be
V^{F}(\mbold{R},\mbold{k}) = g \int \frac{d \mbold{k'}}{(2 \pi)^3}
w(\mbold{k},\mbold{k'}) f_W(\mbold{R},\mbold{k'}) ,
\label{eq23} \ee
where $g$ stands for the degeneracy, $w(\mbold{k},\mbold{k'})$ is
the Fourier transform of the nuclear
interaction $v(\mbold{R},\mbold{R'})$, and
$f_W(\mbold{R},\mbold{k})$ is the distribution function (Wigner
transform of the density matrix).

Following ref. \cite{Cen98}, the distribution function for a non-local
potential with spherical symmetry about $\mbold{k}$ reads
\bea
\tilde{f}_{W}(\mbold{R},k) &=&
\frac{1}{4 \pi} \int f_{WK}(\mbold{R},\mbold{k}) d \Omega_k  \nonumber \\
&=& \Theta (\lambda - \tilde{H}_W - H_{so})
- \frac{\hbar^2}{8 m} \delta'(\lambda - \tilde{H}_W) F_1(\mbold{R},k)
+ \frac{\hbar^2}{24 m} \delta''(\lambda - \tilde{H}_W)
F_2(\mbold{R},k) \nonumber \\
&=& \Theta (\lambda - \tilde{H}_W)
- \frac{\hbar^2}{8 m} \delta'(\lambda - \tilde{H}_W) F_1(\mbold{R},k)
+ \frac{\hbar^2}{24 m} \delta''(\lambda - \tilde{H}_W)
F_2(\mbold{R},k) \nonumber \\ 
&+& \delta(\lambda - \tilde{H}_W)
(\mbold{W} \times
 \mbold{k}) \sigma - \frac{1}{2} \delta'(\lambda - \tilde{H}_W)
(\mbold{W} \times \mbold{k})^2
\label{eq24} \eea
In eq. (\ref{eq24}) the functions $F_1(\mbold{R},k)$ and
$F_2(\mbold{R},k)$ are given by
\be
F_1(\mbold{R},k) = \frac{\hbar^2}{3 m} \big[ \frac{m}{\hbar^2} \Delta V
(3f + k f_k) -k^2 (\bnabla{f})^2 \big]
\label{eq25} \ee
\be
F_2(\mbold{R},k) = \frac{\hbar^2}{3 m} \big[ \frac{m}{\hbar^2} (\bnabla{V})^2
(3f + k f_k) + k^2 f^2 \Delta V -2 k^2 f \bnabla{V} \bnabla{f} \big] ,
\label{eq26} \ee
where $f$ is the inverse of the
position and momentum dependent effective mass:
\be
f(\mbold{R},k) = \frac {m}{m^{*}(\mbold{R},k)}
= 1 + \frac {m}{\hbar^2 k} V^{F}_{k} (\mbold{R},k) ,
\label{eq27} \ee
and the subscript $k$ indicates a partial derivative with respect to
$k$.
The density matrix in coordinate space is given by the inverse Wigner
transform of (\ref{eq25}):
\bea
\rho_{W}(\mbold{R},s) &=& \int \frac{d \mbold{k}}{(2 \pi)^3}
\tilde{f_W}(\mbold{R},s) e^{i \mbold{k} \mbold{s}} \nonumber \\
&=& \frac{k_F^3}{6 \pi^3} \frac{3 j_1(k_F s)}{k_F s} +
\rho_{2,W}(\mbold{R},s,k_F) .
\label{eq28} \eea
The ETF density matrix is obtained from the WK density matrix by
expanding $k_F$
into its $\hbar^0$ and $\hbar^2$ parts and eliminating the spatial
derivatives of the HF potential in favour of the local density and
its gradients \cite{Cen98,SV}. For each kind of nucleon and after
some algebra one finds
\bea
\tilde{\rho}(\mbold{R},s) & = & \frac {3 j_1 (k_F s)}{k_F s} \rho
+ \frac {s^2}{216} \bigg\{ \big[ (9 - 2 k_F \frac {f_k}{f} - 2 k_F^2
\frac {f_{kk}}{f} + k_F^2 \frac {f_k^2}{f^2}) \frac {j_1(k_F s)}{k_F
s}
- 4 j_0(k_F s) \big] \frac{(\bnabla{\rho})^2}{\rho}
\nonumber \\ [2mm]  && \mbox{}
- \big[ (18 + 6 k_F \frac{f_k}{f}) \frac{j_1(k_F s)}{k_F s} -
3 j_0(k_F s) \big] \Delta \rho
\nonumber \\ [2mm] && \mbox{}
- \big[ 18 \rho \frac{\Delta f}{f} + (18 - 6 k_F \frac{f_k}{f})
\frac{\bnabla{\rho}.\bnabla{f}}{f} + 12 k_F
\frac{\bnabla{\rho}.\bnabla{f_k}}{f} - 9 \rho
\frac{(\bnabla{f})^2}{f^2} \big] \frac{j_1(k_F s)}{k_F s} \bigg\}
\nonumber \\ [2mm] && \mbox{}
- \frac{m^2}{\hbar^4} \frac{\rho W^2}{f^2} s^2
\frac{j_1(k_F s)}{k_F s} - \frac{i m}{2 \hbar^2} \frac{\rho}{f}
[\mbold{\sigma} (\mbold{W} \times \mbold{s})] \frac{3 j_1(k_F s)}{k_F
s} ,
\label{eq29} \eea
where now $k_F=(3 \pi^2 \rho)^{1/3}$ and the inverse effective mass
$f$ (\ref{eq25}) and its derivatives with respect to $k$ (subindex
$k$) and $\mbold{R}$ ($\bnabla f$) are computed at $k=k_F$.

The kinetic energy density for each kind of nucleon is obtained from
eq. (\ref{eq17}) using (\ref{eq29}):
\bea
\tau_{ETF}(\mbold{R}) &=& \tau_{ETF,0}(\mbold{R}) + \tau_{ETF,2}
(\mbold{R}) \nonumber \\ &=&
\frac{3}{5} k_F^2 \rho + \frac {1}{36} \frac {(\bnabla{\rho})^2}{\rho}
\big[ 1 + \frac{2}{3} k_F \frac{f_k}{f} + \frac{2}{3} k_F^2
\frac{f_{kk}}{f} - \frac{1}{3} k_F^2 \frac {f_k^2}{f^2} \big] +
\frac{1}{12} \Delta \rho \big[ 4+ \frac{2}{3} k_F \frac{f_k}{f} \big]
\nonumber \\ &+& \frac{1}{6} \rho \frac{\Delta f}{f} + \frac{1}{6}
\frac{\bnabla{\rho}.\bnabla{f}}{f} \big[ 1 - \frac{1}{3} k_F
\frac{f_k}{f} \big] + \frac{1}{9}
\frac{\bnabla{\rho}.\bnabla{f_k}}{f}
- \frac{1}{12} \rho \frac{(\bnabla{f})^2}{f^2} \nonumber \\ &+&
\frac{1}{2} (\frac{2 m}{\hbar^2})^2 \frac{\rho}{f^2} W^2 .
\label{eq30} \eea
The semiclassical spin-orbit density is derived from
(\ref{eq18}) also using (\ref{eq29}):
\bea
\mbold{J}_{ETF}(\mbold{R})&=& -i Tr \{ [\mbold{\sigma} \times
(\frac{\bnabla_R}{2} + \bnabla_s)]
(- \frac{im}{2 \hbar^2}) \frac{\rho}{f} [\mbold{\sigma} (\mbold{W}
\times \mbold{s})] \frac{3 j_1(k_F s)}{k_F s} \} \vert_{s=0} \nonumber
\\
&=& - \frac{3m}{\hbar^2} \frac{\rho}{f} [\bnabla_s \times (\mbold{W}
\times \mbold{s})] \frac{j_1(k_F s)}{k_F s} \vert_{s=0} =
- \frac{2m}{\hbar^2} \frac{\rho \mbold{W}}{f} .
\label{eq31} \eea
The exchange energy density can also be obtained at the ETF level up
to $\hbar^2$ order using the ETF density matrix (\ref{eq29}) and
following the way described in ref. \cite{SV}. It reads
\bea
\varepsilon_{ex}^{ETF}(\mbold{R}) &=&- \frac{1}{2} \int d \mbold{s}
V_{ex}(\mbold{R},s) \rho (\mbold{R} + \frac {\mbold{s}}{2},
\mbold{R} - \frac {\mbold{s}}{2}) \nonumber \\
&=&- \frac{1}{2} \rho (\mbold{R}) \int d \mbold{s}
v(s) \frac{9 j_1^2 ( k_F s)}{k_F^2 s^2} 
+ \frac {\hbar^2}{2 m} \big[ (f - 1)
(\tau_{ETF} - \frac{3}{5} k_F^2 \rho - \frac{1}{4} \Delta \rho)
\nonumber \\
&+& k_F f_k (\frac{1}{27} \frac{(\bnabla{\rho})^2}{\rho}
- \frac{1}{36} \Delta \rho) \big] .
\label{eq32} \eea
The spin-orbit energy density is also easily obtained in the ETF
approach:
\be
\varepsilon^{ETF}_{so}(\mbold{R})= - \frac{2m}{\hbar^2}
\frac{\rho W^2}{f} .
\label{eq33} \ee
Using eqs. (\ref{eq30}), (\ref{eq32}) and (\ref{eq33}) the HF ETF
energy can be written as
\bea
E_{HF}^{ETF} &=& \sum_q \int d \mbold{R} \bigg[ \frac{\hbar^2}{2m}
\frac{3}{5} k_F^2 \rho  + \frac{1}{2} \rho V^{H}
- \frac{1}{2} \rho \int d \mbold{s}
v(s) \frac{9 j_1^2 ( k_F s)}{k_F^2 s^2} \nonumber \\
&+& \frac {\hbar^2}{2 m} \{ f \bigg[
\frac {1}{36} \frac {(\bnabla{\rho})^2}{\rho}
\big[ 1 + \frac{2}{3} k_F \frac{f_k}{f} + \frac{2}{3} k_F^2
\frac{f_{kk}}{f} - \frac{1}{3} k_F^2 \frac {f_k^2}{f^2} \big] +
\frac{1}{12} \Delta \rho \big[ 1+ \frac{2}{3} k_F \frac{f_k}{f} \big]
\nonumber \\ &+& \frac{1}{6} \rho \frac{\Delta f}{f} + \frac{1}{6}
\frac{\bnabla{\rho}.\bnabla{f}}{f} \big[ 1 - \frac{1}{3} k_F
\frac{f_k}{f} \big] + \frac{1}{9}
\frac{\bnabla{\rho}.\bnabla{f_k}}{f}
- \frac{1}{12} \rho \frac{(\bnabla{f})^2}{f^2} \bigg] \nonumber \\
&+& \frac{1}{4} \Delta \rho
+ k_F f_k (\frac{1}{27} \frac{(\bnabla{\rho})^2}{\rho} - \frac{1}{36}
\Delta \rho) \}
- \frac{1}{2} \frac{2 m}{\hbar^2} \frac{\rho}{f} W^2 \bigg]_q .
\label{eq34} \eea
Now the variational equations read
\be
\frac{\delta}{\delta \rho_n} \big[ E^{ETF}_{HF} - \mu_n \int
\rho_n (\mbold{R}) d \mbold{R} \big] = 0
\label{eq35} \ee
\be
\frac{\delta}{\delta \rho_p} \big[ E^{ETF}_{HF} - \mu_p \int
\rho_p (\mbold{R}) d \mbold{R} \big] = 0 .
\label{eq36} \ee
 This is a set of two coupled second-order non-linear
differential equations that can be solved using, for instance, the
imaginary time-step method \cite{Cen90} and allows one to find the
semiclassical densities $\rho_n(\mbold{R})$ and
$\rho_p(\mbold{R})$ which are the fully variational solutions
of the ETF HF energy (\ref{eq34}).

\section{Skyrme Forces}

The Skyrme forces \cite{Sky} are among the most important and most
widely used phenomenological nuclear forces due to their simplicity
because of their zero range. The Skyrme force consists of some
two-body
terms together with a three-body term that can be replaced by a
density dependent two-body contribution:
\bea
v(\mbold{s}) &=& t_0(1+x_0 P^{\sigma}) \delta(\mbold{s})
+ \frac{1}{2} t_1(1+x_1 P^{\sigma}) [ \delta(\mbold{s})
\hat{\mbold{k}}^2
+ \hat{\mbold{k'}}^2 \delta(\mbold{s}) ] + t_2(1+x_2 P^{\sigma})
\hat{\mbold{k'}} \delta(\mbold{s}) \hat{\mbold{k}}
\nonumber \\ &+& \frac{1}{6}t_3(1+x_3 P^{\sigma})
[\rho(\mbold{R})]^{\alpha} \delta(\mbold{s})
+ iW_0 (\mbold{\sigma}_1 + \mbold{\sigma}_2) \hat{\mbold{k'}}
\times \delta(\mbold{s}) \hat{\mbold{k}} ,
\label{eq37} \eea
where $\mbold{R}$ and $\mbold{s}$ are the center of mass and relative
coordinates. The relative momentum operators $\hat{\mbold{k}}$ and
$\hat{\mbold{k'}}$ act on the right and on the left, respectively.

The ground-state Skyrme HF energy can be written in terms of an
integral of the energy density which has the following structure:
\bea
\varepsilon(\mbold{R}) &=& \frac{\hbar^2}{2m} (\tau_n + \tau_p)
\nonumber \\
&+& \frac{1}{2} \rho^2 [t_0(1+\frac{x_0}{2}) + \frac{1}{6}t_3
\rho^{\alpha}(1+\frac{x_3}{2})]
- \frac{1}{2}(\rho_n^2+\rho_p^2)
[t_0(x_0+\frac{1}{2}) + \frac{1}{6}t_3 \rho^{\alpha}(x_3+\frac{1}{2})]
\nonumber \\ &+& \frac{1}{4} \rho \tau [t_1(1+\frac{x_1}{2}) +
t_2(1+\frac{x_2}{2})] -\frac{1}{4}(\rho_n \tau_n + \rho_p \tau_p)
[t_1(x_1+\frac{1}{2})-t_2(x_2+\frac{1}{2})] \nonumber \\ &+&
\frac{1}{16}
(\bnabla{\rho})^2 [3t_1(1+\frac{x_1}{2})-t_2(1+\frac{x_2}{2})]
\nonumber \\ &-&
\frac{1}{16}((\bnabla{\rho_n})^2 + (\bnabla{\rho_p})^2)
[3t_1(x_1+\frac{1}{2})+t_2(x_2+\frac{1}{2})] \nonumber \\ &+&
\frac{1}{2}W_0[ \mbold{J} \bnabla{\rho} + \mbold{J}_n
\bnabla{\rho_n} + \mbold{J}_p \bnabla{\rho_p}] ,
\label{eq38} \eea
where the particle, kinetic energy and spin-orbit densities are given
by eqs. (\ref{eq16})-(\ref{eq18}).

The variation of the ground-state energy with respect to the
single-particle wave functions $\phi_i^{*}$ leads to the following
set of HF equations:
\be
\{ -\bnabla \frac{1}{2m_q^{*}(\mbold{r})}\bnabla + U_q(\mbold{r})
- i \mbold{W}(\mbold{r}) (\bnabla \times \mbold{\sigma})\}
\phi_{i,q}(\mbold{r}) = \epsilon_{i,q} \phi_{i,q} (\mbold{r}) .
\label{eq39} \ee
The local potential,
the effective mass $m_q^{*}$ and the spin-orbit potential
$\mbold{W}(\mbold{r})$ are given by
\bea
U_q(\mbold{r})&=& \frac{\delta \varepsilon}{\delta \rho_q} =
 \rho [t_0(1+\frac{x_0}{2}) + \frac{1}{12}(\alpha+2)
\rho^{\alpha}(1+\frac{x_3}{2})]
- \rho_q [t_0(1+\frac{x_0}{2}) + \frac{1}{6}t_3
\rho^{\alpha}(1+\frac{x_3}{2})] \nonumber \\ &-& \frac{1}{12}[
\rho_n^2+\rho_p^2]
t_3 \alpha \rho^{\alpha-1}(1+\frac{x_3}{2})
\nonumber \\ &+& \frac{1}{4} \tau [t_1(1+\frac{x_1}{2}) +
t_2(1+\frac{x_2}{2})] -\frac{1}{4} \tau_q
[t_1(x_1+\frac{1}{2})-t_2(x_2+\frac{1}{2})] \nonumber \\ &-&
\frac{1}{8}
\Delta \rho [3t_1(1+\frac{x_1}{2})-t_2(1+\frac{x_2}{2})]
+\frac{1}{8} \Delta \rho_q
[3t_1(x_1+\frac{1}{2})+t_2(x_2+\frac{1}{2})] \nonumber \\ &-&
\frac{1}{2}W_0[ \bnabla \mbold{J} + \bnabla \mbold{J}_q] ,
\label{eq40} \eea
\be
\frac{\hbar^2}{2 m_q^{*}(\mbold{r})} = \frac{\delta
\varepsilon}{\delta \tau_q} = 
\frac{\hbar^2}{2m} + \frac{1}{4} \rho  [t_1(1+\frac{x_1}{2}) +
t_2(1+\frac{x_2}{2})] - \frac{1}{4} \rho_q
[t_1(x_1+\frac{1}{2})-t_2(x_2+\frac{1}{2})] ,
\label{eq41} \ee
\be
\mbold{W}_q(\mbold{r}) = \frac{\delta \varepsilon}{\delta \mbold{J}_q}
= \frac{1}{2} W_0 [ \bnabla \rho + \bnabla \rho_q] .
\label{eq42} \ee
Notice that for Skyrme forces the HF theory coincides with the
Kohn-Sham theory extended for effective mass and spin-orbit
contributions
\cite{Brack,Bra1}. This is due to the fact that in this case the full
potential energy density can be written as a functional of the local
density, see eq. (\ref{eq40}). From the point of view of the Kohn-Sham
scheme,
correlations beyond HF are also included. In the present case of the
Skyrme
forces, they are implicitly contained in the parameters which are
fitted to reproduce the experimental data.

To apply the ETF approach in the way described in Section 2, notice
that for Skyrme forces the Wigner transform of eq. (\ref{eq12}) reads
\be
H_W = \frac{\hbar^2 k^2}{2m} + U_q(\mbold{R}) +
\frac{\hbar^2 k^2}{2m}(f_q(\mbold{R})-1) + \frac{\hbar^2}{8m} \Delta
f_q(\mbold{R})
- (\mbold{W}(\mbold{R}) \times \mbold{k}) \mbold{\sigma}
\label{eq43} \ee
where in this case the effective mass $f_q$ is only position
dependent, see eq. (\ref{eq41}).
However, in eq. (\ref{eq43}), $U_q$ and $\hbar^2 k^2 (f_q-1)/2m$ do not
correspond to the Hartree and Fock potentials. These two terms are
obtained as functional derivatives of the energy density (\ref{eq38})
that contains both direct and exchange contributions. For Skyrme
forces the $k$-dependence in $H_W$ comes from the explicit dependence
on the relative momentum operator of the interaction (\ref{eq37})
that contributes to the Hartree and Fock parts of the single-particle
potential.

For Skyrme forces, the kinetic energy density found from
(\ref{eq30}) is given by
\be
\tau_{ETF}(\mbold{R}) = \frac{3}{5} k_F^2 \rho + \frac{1}{36}
\frac{(\bnabla{\rho})^2}{\rho} + \frac{1}{3} \Delta \rho
+\frac{1}{6} \rho \frac{\Delta f}{f} + \frac{1}{6}
\frac{\bnabla{\rho}.\bnabla{f}}{f} - \frac{1}{12} \rho
\frac{(\bnabla{f})^2}{f^2} + \frac{1}{2}
\big(\frac{2m}{\hbar^2}\big)^2 \frac{\rho}{f^2} \mbold{W}^2
\label{eq44} \ee
in accordance with the result of ref. \cite{Brack}.
The ETF energy density given by (\ref{eq38}) with the kinetic energy
and spin-orbit densities replaced by their semiclassical counterparts
eqs. (\ref{eq44}) and (\ref{eq31}) respectively. In this way the HF
ETF energy density becomes a functional of the proton and neutron
densities that are obtained by solving the corresponding
Euler-Lagrange equations (\ref{eq35}) and (\ref{eq36}).

As it has been discussed in the introduction, the semiclassical ETF
energy should be similar to the one obtained using the more
complicated Strutinsky average. However, as it has been pointed out in
previous literature \cite{Bra1}, if the ETF kinetic energy density is
calculated to $\hbar^2$ order only, its integral is not able to
reproduce the Strutinsky kinetic energy, at least in the case of a set
of nucleons moving in a harmonic oscillator or a Woods-Saxon external
potential. Consequently, $\hbar^4$ contributions to the kinetic
energy and spin-orbit densities have to be taken into account. We will
not give the explicit expressions of these functionals $\tau_4[\rho]$
and $\mbold{J_4}[\rho]$ that can be found for instance in ref.
\cite{Gram} in the case of single-particle
Hamiltonians whose Wigner transform is of the type (\ref{eq43}).

The Euler-Lagrange equations associated to the ETF HF energy including
$\hbar^4$ corrections were solved by the first time in
ref. \cite{Cen90},
where a detailed discussion of the semiclassical kinetic and
spin-orbit energies can be found.
We now present variational semiclassical ETF-$\hbar^4$ results for
binding energies, densities and radii of some selected double magic
nuclei and compare them with the fully quantal results. In these
applications we have used the recently presented SLy4 \cite{Cha}
parametrization of the Skyrme force, which is able to describe nuclei
far from
the stability lines. The results for binding energies and radii are
collected in Table 1.

From Table 1 we can see that the semiclassical ETF energies of
$\hbar^4$ order are close to the HF ones pointing out that,
according the Strutinsky energy theorem, the shell energy is small and
can be added perturbatively.
This can be done performing a Strutinsky calculation using the
semiclassical $U_q$, $m_q^{*}$ and $\mbold{W}_q$ Skyrme mean fields.
Another alternative is to use the so-called expectation value
method (EVM)
\cite{Brack,BCKT} that consists in performing one HF iteration using
the semiclassical mean fields as input. The binding energies obtained
using the EVM are also collected in Table 1. They are smaller than the
HF energies, in accordance with the Ritz variational principle, by
less than 3 MeV in all the considered nuclei. This shows that the
EVM allows one to
obtain rather accurate total binding energies including shell effects
at the cost of essentially one microscopic HF step beyond the
semiclassical calculation.

As a further illustration, Figure 1 displays the quantal HF
particle and kinetic energy densities for neutrons and protons
calculated in $^{208}$Pb and $^{132}$Sn 
with SLy4, as well as the results of the
ETF-$\hbar^4$ and EVM approximations. Figure 2 is a complementary plot
showing the densities for $^{208}$Pb in the outer
surface region on a semi-logarithmic scale. One can compare the
exponential fall-off of the HF and EVM densities with the $1/r^6$
behaviour of the ETF-$\hbar^4$ solution at large distances.

\section{Finite range forces}

Now we turn our attention to the discussion of finite range forces
as applied
to semiclassical HF calculations. We will consider here an effective
nucleon-nucleon force of Gogny type \cite{Gog}. It consists 
of a central finite range part together with
zero-range density-dependent and spin-orbit contributions:
\bea
v(s)&=&\sum_{i=1}^2 [(w_i+\frac{b_i}{2}P^{\sigma}-\frac{h_i}{2}P^{\tau}-
\frac{m_i}{4}P^{\sigma}P^{\tau})\tilde{w}_i(s)] \nonumber \\
&+& \frac{1}{6}t_3(1+x_3 P^{\sigma})
[\rho(\mbold{R})]^{\alpha} \delta(\mbold{s})
+ iW_0 (\mbold{\sigma}_1 + \mbold{\sigma}_2) \hat{\mbold{k'}}
\times \delta(\mbold{s}) \hat{\mbold{k}} , \label{eq45}
\eea
where $w_i$, $m_i$, $b_i$ and $h_i$ are the usual exchange parameters
of the central force
and $\tilde{w}_i(s)=\exp(-s^2/\mu_i^2)$ are the Gaussian form factors.
The contributions of the zero-range and spin-orbit parts of
(\ref{eq45})
to the HF energy and single-particle potential (SPP) are the same as for
the Skyrme force described in the previous section. Therefore, in the
following we will concentrate only on the contributions to the energy
and SPP associated with the finite range part of (\ref{eq45}).

In the ETF approximation the HF energy is given by eq. (\ref{eq34})
which yields the SPP potential through the variational principle.
For the Gogny force the direct potential is given by \cite{Soub}
\bea
V_q^H(\mbold{R})&=& \sum_{i=1}^2 \frac{\pi \mu_i^2}{R}
\int_{0}^{\infty} dr r [\exp(-\frac{(R-r)^2}{\mu_i^2}) -
\exp(-\frac{(R+r)^2}
{\mu_i^2})] \nonumber \\ && 
\mbox{} \times [(w_i+\frac{b_i}{2}) \rho(r) - (h_i
+\frac{m_i}{2}) \rho_q(r)] .
\label{eq46} \eea
The $\hbar^0$ order exchange energy is given by \cite{Soub}
\be
\varepsilon_{ex}^{0}(\mbold{R}) = - \sum_{i=1}^{2} \frac{1}{6 \pi^2 \mu_i^3}
\{ X_{e,1,i} [v_i(k_{Fn},k_{Fn}) + v_i(k_{Fp},k_{Fp})] - 2 X_{e,2,i}
v_i(k_{Fn},k_{Fq})\} ,
\label{eq47} \ee
where $X_{e,i,1}=w_i/2+b_i-h_i/2-m_i$ and
$X_{e,2,i}=h_i/2+m_i$. The functions $v_i(k_{Fq},k_{Fq'})$ read
\bea
v_i(k_{Fq},k_{Fq'}) &=& \mu_i^3(k_{Fq}+k_{Fq'})(k_{Fq}^2+k_{Fq'}^2-k_{Fq}
k_{Fq'}) {\rm erf} [\frac{\mu_i(k_{Fq}+k_{Fq'})}{2}] \nonumber \\
&-& \mu_i^3(k_{Fq}-k_{Fq'})(k_{Fq}^2+k_{Fq'}^2+k_{Fq}k_{Fq'})
 {\rm erf} [\frac{\mu_i(k_{Fq}-k_{Fq'})}{2}] \nonumber \\
&+& \frac{2}{\sqrt{\pi}}[\mu_i^2(k_{Fq}^2+k_{Fq'}^2-k_{Fq}k_{Fq'}) -2]
\exp[- \frac{\mu_i^2(k_{Fq}+k_{Fq'})^2}{4}] \nonumber \\
&-& \frac{2}{\sqrt{\pi}}[\mu_i^2(k_{Fq}^2+k_{Fq'}^2+k_{Fq}k_{Fq'}) -2]
\exp[- \frac{\mu_i^2(k_{Fq}-k_{Fq'})^2}{4}] .
\label{eq48} \eea
The $\hbar^0$ order exchange potential in phase space is
\be
V_q^F(\mbold{R},k) = - \sum_{i=1}^2 [X_{e,1,i} u_i(k,k_{Fq})
-X_{e,2,i} u_i(k,k_{Fq'})] ,
\label{eq49} \ee
where the functions $u_i(k,k_{Fq})$ read
\bea
u_i(k,k_{Fq}) &=&
{\rm erf} [\frac{\mu_i(k+k_{Fq})}{2}]
-{\rm erf} [\frac{\mu_i(k-k_{Fq})}{2}]
\nonumber \\
&+& \frac{2}{\sqrt{\pi} \mu_i k}
( \exp[- \frac{\mu_i^2(k+k_{Fq})^2}{4}]
- \exp[- \frac{\mu_i^2(k-k_{Fq})^2}{4}] ) .
\label{eq50} \eea
It follows that the inverse of the momentum and position dependent
effective mass
(\ref{eq27}) needed to compute the $\hbar^2$ contributions to the
kinetic and exchange energy densities is 
\be
f_q(\mbold{R},k) = 1 + \frac{m}{\hbar^2 k} \sum_{i=1}^2 [X_{e,1,i}
\frac{\partial u_i(k,k_{Fq})}{\partial k}-X_{e,2,i}
\frac{\partial u_i(k,k_{Fq'})}{\partial k}] .
\label{eq51} \ee

With the help of eqs. (\ref{eq46}), (\ref{eq47}) and (\ref{eq51}) the
full energy density (\ref{eq34}) plus the $t_3$ contribution of
(\ref{eq38}) can be written. Next, from this functional one should
derive the ETF variational equations.
This implies some lengthy algebra that can be partially avoided as
follows. As shown in ref. \cite{SV}, one obtains almost the
same ETF-$\hbar^2$ energy if the full one-body semiclassical density
matrix (\ref{eq29}) is replaced by the one corresponding to a local
potential (i.e., dropping all the space and momentum derivatives of
$f$ in eq. (\ref{eq29}) for the density matrix).
With this simplification the kinetic energy density (\ref{eq30}),
which also appears in the $\hbar^2$ contribution to the exchange
energy density, reduces to the one corresponding to a local potential.
We use here this approximate way for deriving the variational equations.
Once these equations have been solved, we compute the energy using the
complete expression of the energy density.

The semiclassical ETF-$\hbar^2$ binding energies and r.m.s. radii 
obtained with the Gogny force D1 for some selected magic nuclei are
reported in Table 2, which compares
them with the quantal HF values. Usually, HF calculations with Gogny
forces are carried out taking into account the two-body part of the
center-of-mass correction. In the semiclassical framework one finds
that the two-body correction exactly cancels the one-body part
\cite{Brack}. Thus, we have not included the center-of-mass correction
in the semiclassical results presented in Table 2. It is also known
that the ETF approximation at order $\hbar^2$ overbinds the nuclei and
gives smaller r.m.s. radii than the HF ones \cite{Brack,Cen90,BCKT}.
These trends are followed in general by the semiclassical results with
the D1 force as seen from Table 2.

One possible way to recover quantal effects, which are absent in the
ETF approach described above, consists in considering the $\hbar^2$
exchange energy density as the exchange-correlation energy density in
the Kohn-Sham scheme, and solving for each single-particle state the
corresponding local Schr\"odinger equation (\ref{eq7}). For this
purpose, we replace $\rho$ and $\tau$ in the semiclassical exchange
energy density (\ref{eq32}) by the Kohn-Sham ansatz given by  eqs.\
(\ref{eq5}) and (\ref{eq6}), which allows us to write the
corresponding KS equations including effective mass and spin-orbit
contributions that are similar to the HF equations for Skyrme forces.

To illustrate this approach we report in Table 3 our KS binding
energies for tin isotopes in comparison with the HF values given in
ref. \cite{Gog}, which do not include the two-body center-of-mass
correction. We realize that our KS-$\hbar^2$ binding energies nicely
reproduce the HF energies, the discrepancies being less than 2 MeV in
this region of tin isotopes. For comparison, we also present in Table
3 the KS-$\hbar^0$ results  from ref. \cite{Soub}
(where only the Slater term of the binding
energy is taken as exchange-correlation energy).
We can see that taking into account the $\hbar^2$-order corrections in
our KS approach clearly improves the KS-$\hbar^0$ results.

%
\section*{Conclusions}

In this paper we have reviewed the ETF approach to the HF method. From
a theoretical point of view, we have derived the semiclassical binding
energy for a non-local potential related with a finite range effective
interaction, starting from the corresponding ETF density matrix up to
order $\hbar^2$. As a limiting case, one recovers for Skyrme forces
the usual expression for the kinetic energy density.

As a first numerical example, we have presented some ETF calculations
of binding energies and r.m.s. radii of some magic nuclei using a
recently proposed parametrization of the Skyrme force (SLy4). These
semiclassical calculations have been carried out including corrections
of order $\hbar^4$. These ETF binding energies are close to the HF
ones and are similar to the ones obtained using a Strutinsky smoothing
procedure. To recover the shell effects absent in the semiclassical
calculation, one can use the expectation value method. It basically
consists in performing one quantal iteration on top of the
semiclassical calculation. In this way one reproduces the HF results
with an accuracy around 0.5\% for all the studied nuclei.

Finally, we have performed semiclassical ETF calculations of binding
energies to order $\hbar^2$ using the finite range D1 Gogny force.
These energies overbind the HF results by an amount around 3.5\%, as
it happens in the case of zero range Skyrme forces if the
semiclassical calculation is taken only to second order in
$\hbar$. We have also used the ETF-$\hbar^2$ exchange energy density
as an exchange-correlation energy density within the Kohn-Sham scheme.
In this
case, we find that the KS binding energies reproduce almost perfectly
the HF energies in the region of tin isotopes studied.

%
\section*{Acknowledgements}

Two of us (X.V. and M.C.) would like to acknowledge support from the DGICYT (Spain)
under grant PB95-1249 and from the DGR (Catalonia) under grant
1998SGR-00011. The support of the agreement of the St. Petersburg and 
Barcelona Universities is
also acknowledged.

%

\pagebreak
%
\section*{Table 1}

The quantal HF binding energies (in MeV) and r.m.s.\ neutron and
proton radii
(in fm) of double closed shell nuclei for the Skyrme interaction SLy4
are compared with the results of the semiclassical $\hbar^4$-order
ETF calculation and of the EVM discussed in the text.

\vspace{2cm}
\begin{center}
\begin{tabular}{llcccccccc}
\hline
 & &  $^{16}$O  &  $^{40}$Ca &  $^{48}$Ca &  $^{56}$Ni
   &  $^{78}$Ni & $^{100}$Sn & $^{132}$Sn & $^{208}$Pb \\
\hline
$B$: & HF  &
128.3 & 344.2 & 417.9 & 483.4 & 643.9 & 828.8 & 1103.8 & 1636.1 \\ 
     & EVM &
127.6 & 342.9 & 416.3 & 480.4 & 642.2 & 825.8 & 1102.0 & 1634.3 \\
     & ETF-$\hbar^4$ &
128.7 & 348.8 & 422.8 & 488.1 & 645.5 & 826.9 & 1098.4 & 1629.9 \\
$r_{\rm n}$: & HF  &
2.71 & 3.39 & 3.63 & 3.67 & 4.22 & 4.36 & 4.90 & 5.63 \\
             & EVM &
2.66 & 3.34 & 3.63 & 3.70 & 4.25 & 4.38 & 4.92 & 5.64 \\
             & ETF-$\hbar^4$ &
2.66 & 3.34 & 3.61 & 3.67 & 4.25 & 4.36 & 4.91 & 5.62 \\
$r_{\rm p}$: & HF  &
2.74 & 3.44 & 3.47 & 3.72 & 3.93 & 4.44 & 4.68 & 5.46 \\
             & EVM &
2.69 & 3.39 & 3.45 & 3.75 & 3.96 & 4.46 & 4.70 & 5.48 \\
             & ETF-$\hbar^4$ &
2.68 & 3.38 & 3.46 & 3.72 & 3.94 & 4.44 & 4.69 & 5.48 \\
\hline
\end{tabular}
\end{center}

\pagebreak
\section*{Table 2}

Binding energies (in MeV) and r.m.s.\ neutron and proton radii (in fm)
for the D1 Gogny interaction. The HF binding energies are from ref.
\cite{Mey86} and the HF radii from ref. \cite{Gog}.

\vspace{2cm}
\begin{center}
\begin{tabular}{llcccccccc}
\hline
 & & $^{16}$O & $^{40}$Ca & $^{48}$Ca & $^{90}$Zr & $^{208}$Pb \\
\hline
$B$: & HF  &
127  & 338  & 411  & 779  & 1633 \\
     & ETF-$\hbar^2$ &
123.9 & 350.0 & 424.8 & 807.3 & 1670.9 \\
$r_{\rm n}$: & HF &
2.63 & 3.34 & 3.38 & 4.24 & 5.53 \\
             & ETF-$\hbar^2$ &
2.57 & 3.28 & 3.54 & 4.22 & 5.53 \\
$r_{\rm p}$: & HF &
2.65 & 3.38 & 3.38 & 4.18 & 5.40 \\
             & ETF-$\hbar^2$ &
2.58 & 3.32 & 3.43 & 4.18 & 5.44 \\
\hline
\end{tabular}
\end{center}

\pagebreak
\section*{Table 3}

Binding energies (in MeV) of tin isotopes in the KS-$\hbar^0$,
KS-$\hbar^2$ and HF calculations for the D1 Gogny interaction. The
KS-$\hbar^0$ values are from ref. \cite{Soub} and the HF values are
from ref. \cite{Gog}.

\vspace{2cm}
\begin{center}
\begin{tabular}{lcrcrcr}
\hline
 & & KS-$\hbar^0$ & & KS-$\hbar^2$ & & HF \\
\hline
$^{112}$Sn & &  940.71 & &  949.42 & &  948.30 \\
$^{114}$Sn & &  958.48 & &  969.56 & &  968.40 \\
$^{116}$Sn & &  976.02 & &  986.96 & &  985.70 \\
$^{118}$Sn & &  991.81 & & 1003.27 & & 1002.16 \\
$^{120}$Sn & & 1007.40 & & 1019.77 & & 1018.90 \\
$^{122}$Sn & & 1022.81 & & 1033.30 & & 1032.19 \\
$^{124}$Sn & & 1038.04 & & 1047.10 & & 1045.90 \\
$^{132}$Sn & & 1098.22 & & 1105.03 & & 1102.77 \\
\hline
\end{tabular}
\end{center}

\pagebreak
%

\begin{figure}[t]
\centering
\includegraphics[width=16cm]{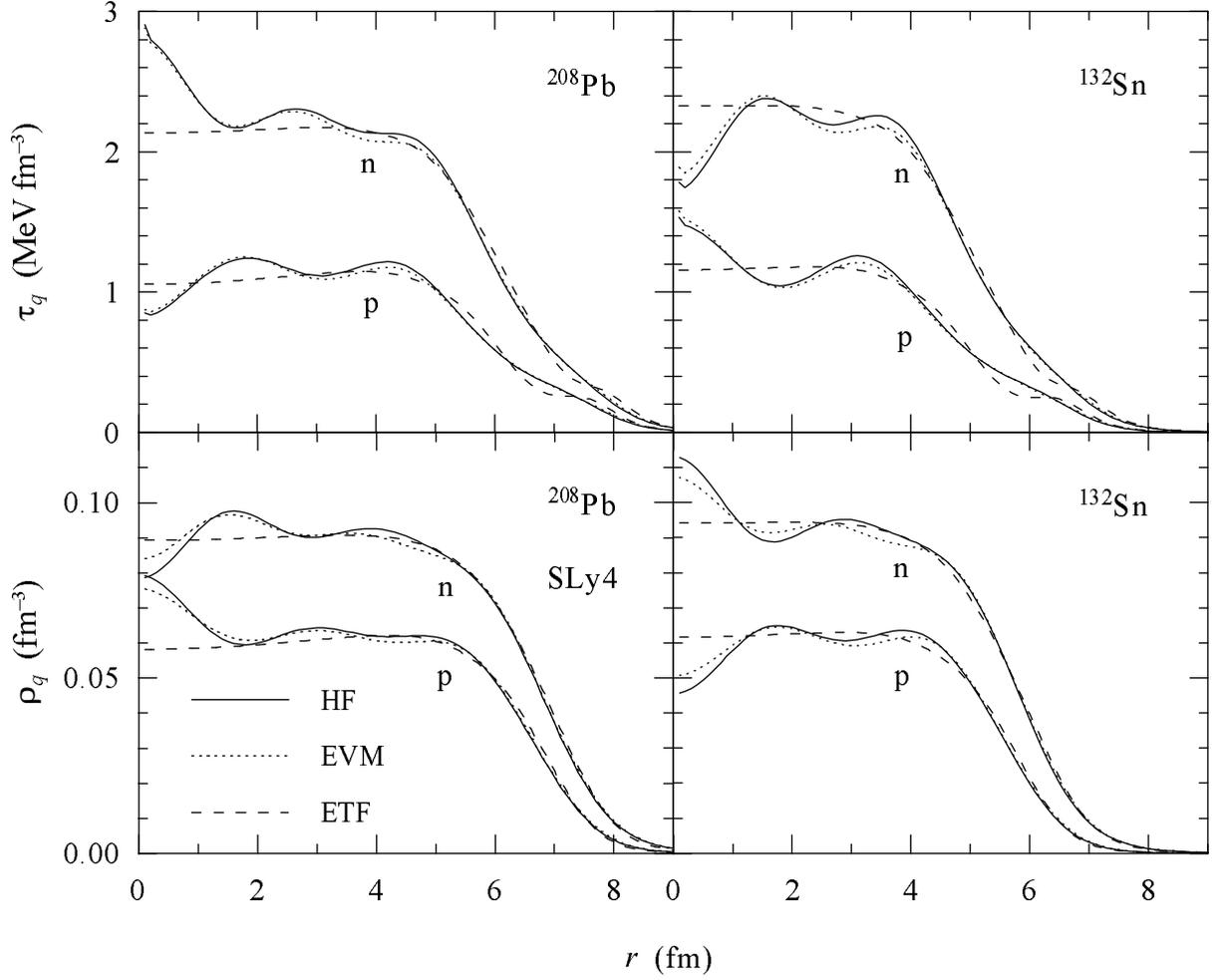}
\caption{Comparison of the particle and kinetic energy densities for neutrons
and protons in $^{208}$Pb and $^{132}$Sn obtained with SLy4 in the HF,
$\hbar^4$-order ETF and EVM approximations.}
\end{figure}

\pagebreak

\begin{figure}[t]
\centering
\includegraphics[width=10cm]{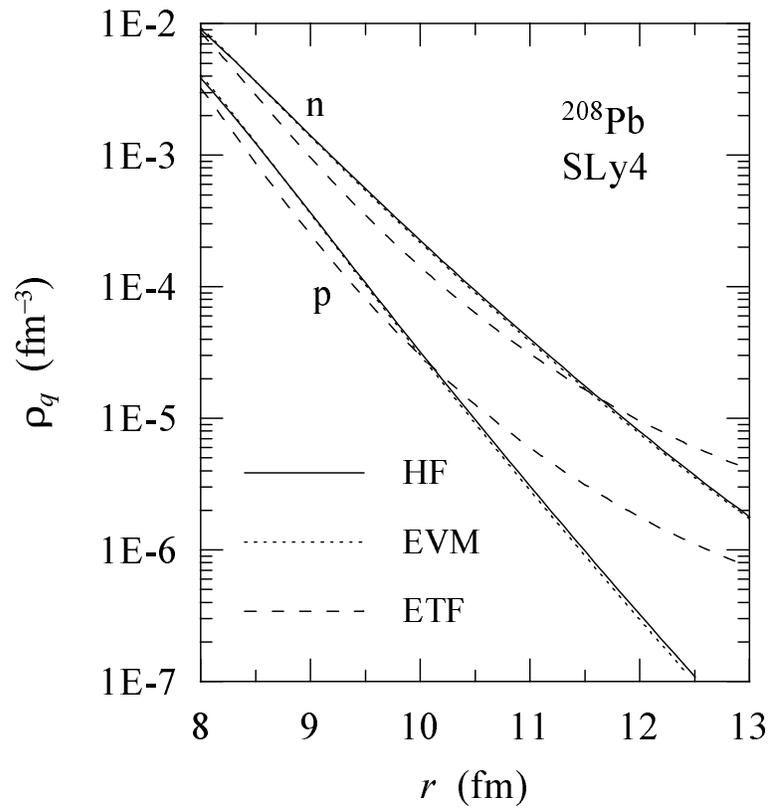}
\caption{Semi-logarithmic plot of the densities of Figure 1 for $^{208}$Pb
in the outer surface region.}
\end{figure}

%
\end{document}